\documentclass[useAMS]{mn2e}
\usepackage{graphicx}

\title[Near-infrared Observations of Nova V574~Pup]{Near-infrared Observations of Nova V574~Puppis (2004)} 
\author[Naik et al.]{Sachindra Naik\thanks{snaik@prl.res.in}, D. P. K. Banerjee\thanks{orion@prl.res.in}, N. M. Ashok\thanks{ashok@prl.res.in},   \& R. K. Das\thanks{rkdas@prl.res.in},
\\
Astronomy and Astrophysics Division, Physical Research Laboratory, Navrangapura, Ahmedabad - 380009, Gujarat, India}
\begin{document}

\date{Accepted for publication in MNRAS}

\maketitle

\begin{abstract}
We present results obtained from  extensive near-infrared spectroscopic
and photometric observations of nova V574~Pup  during its 2004 outburst. 
The observations were obtained over four months, starting from 2004 
November 25 (four days after the nova outburst) to 2005 March 20. The 
near-IR $JHK$ light curve is presented - no evidence is seen from it
for dust formation to have occurred during our observations. In the early 
decline phase, the $JHK$ spectra of the nova are dominated by  emission 
lines of hydrogen Brackett and Paschen series, OI, CI and HeI.  We also 
detect the fairly uncommon Fe~II line at 1.6872 \micron~ in the early 
part of our observations. The strengths of the  HeI lines at 1.0830 
\micron~ and 2.0585 \micron~ are found to become very strong towards 
the end of the observations indicating a progression towards higher 
excitation conditions in the nova ejecta. The width of the emission 
lines do not show any significant change during the course of our 
observations.  The slope of the continuum spectrum was found to have 
a $\lambda^{-2.75}$ dependence in the early stages which gradually 
becomes flatter with time and changes to a free-free spectral dependence 
towards the later stages. Recombination analysis of the HI lines shows 
deviations from Case B conditions during the initial stages. However, 
towards the end of our observations, the line strengths are well 
simulated with case~B model values with electron density $n_e$ = 
10$^{9-10}$ cm$^{-3}$ and a temperature equal to 10$^4$ K. Based on our 
distance estimate to the nova of 5.5 kpc and the observed free-free 
continuum emission in the later part of the observations, we estimate 
the ionized mass of the ejecta to be between 10$^{-5}$ M$_\odot$ and 
10$^{-6}$ M$_\odot$. 
\end{abstract}

\begin{keywords}
infrared: stars -- novae, cataclysmic variables -- stars: individual 
(V574~Pup)-- techniques: spectroscopic
\end{keywords}

\section{Introduction}
V574~Pup was independently discovered to be in outburst by Tago (Nakano 
et al. 2004) on 2004 November 20.67 UT at V$\sim$7.6 and, within a short 
time thereafter, by Sakurai on 2004 November 20.812 UT at V$\sim$7.4 
(Nakano et al. 2004). The follow-up observations reported by Samus \& 
Kazarivets (2004) showed a post-discovery brightening before the onset of
fading. Low dispersion optical spectra of the nova on 2004 November 21.75 
UT showed H$\alpha$ and H$\beta$ emission lines with P~Cygni components, 
along with the strong Fe~II (multiplet 42) in absorption indicating that 
V574~Pup is a ``Fe II'' class nova near maximum light (Ayani 2004). 
Subsequent near-infrared spectroscopic observations of the nova on 2004 
November 26.98 UT showed strong HI emission lines from the Paschen and 
Brackett series, OI lines at 1.1287 \& 1.3164 \micron~ and a blend of 
NI \& CI lines in the spectral region 1.175 to 1.25 \micron~ (Ashok 
\& Banerjee 2004). The optical spectra of the nova on 2004 November 
26 and December 12 obtained by Siviero et al. (2005) were found to be 
dominated by Balmer hydrogen and FeII emission lines; no nebular lines 
were present in the spectra of December 12. One year after the outburst, 
the nova was found to be well into the coronal phase with the detection 
of [Si~VI], [Si~VII], [Ca~VIII], [S~VIII], and [S~IX] lines in its 
spectrum (Rudy et al. 2005). Along with these lines, unidentified nova 
features at 0.8926, 1.1110, 1.5545 and 2.0996 \micron~ were also present 
in the spectrum. However, no evidence for emission from dust was seen in 
these observations.
 
V574~Pup was observed by the Spitzer Space Observatory one year after 
the outburst revealing strong coronal lines (Rudy et al. 2006). The 
spectroscopic observations by Lynch et al. (2007), three years after 
the outburst, showed the persistence of the coronal phase. Rudy et al. 
(2006) have remarked that the presence of strong coronal emission lines 
suggests similarity with a He/N nova. Siviero et al. (2005) have inferred 
that V574~Pup suffers negligible interstellar extinction though it is 
close to the galactic plane (b=-2$^\circ$). They estimate that the nova 
is  located at very large distance of 15 to 20 kpc. In a later subsection 
we determine the distance to V574~Pup using the optical light 
curve and MMRD relation. V574~Pup is one of the  novae detected with X-ray 
emission, among 12 classical novae  studied by Ness et al. (2007) using 
Swift observations. It was observed on several occasions between 2005 May 
and 2005 August by Swift. The X-ray spectra showed it to be in the super 
soft X-ray  phase. Ness et al. (2007) estimate the color excess $E(B-V)$ = 
0.5 and the distance to be 3.2 kpc.

\begin{figure}
\vskip 7.0 cm
\includegraphics{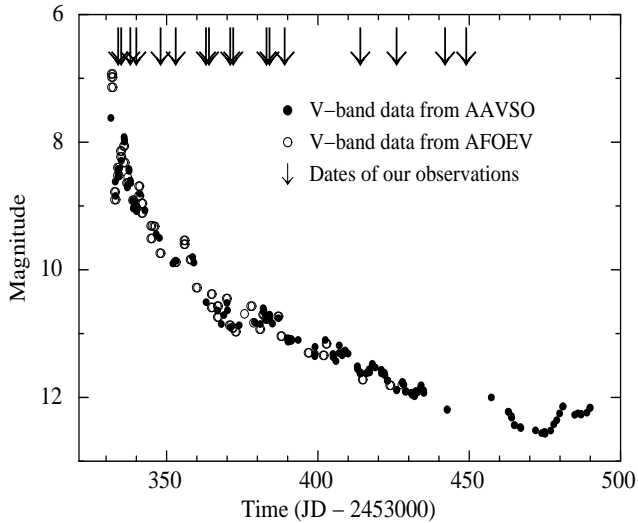}
\caption{The V-band light curve of V574~Pup obtained from the 
AAVSO (shown by filled circles) and AFOEV data (shown by open circles). 
The arrow marks show the days of our near-infrared observations.}
\end{figure}

\section{Observations and Data Reduction}
Near-infrared spectroscopic and photometric observations of V574~Pup 
were carried out fairly extensively using the 1.2-m telescope of the Mt Abu 
Infrared Observatory. The $V$ band light curve is shown in Figure~1 
with the epochs of our observations  marked by arrows -  the log of the 
observations is given in Table~1. The Mt. Abu spectra were obtained at a 
resolution of $\sim$1000 using a Near-Infrared Imager/Spectrometer with a 
256$\times$256 HgCdTe (NICMOS3) array. Photometric observations of the nova 
were carried out on several nights (Table~1) in photometric sky conditions 
using the NICMOS3 array in the imaging mode. Several frames were obtained 
in four dithered positions, typically offset by $\sim$ 30 arcsec from each 
other, with exposure times ranging from 0.4--100 s depending on the 
brightness of the nova. The sky frames were generated by median combining 
the average of each set of dithered frames and  subsequently subtracted 
from the nova frames. A nearby field-star SAO~174367, observed at similar 
airmass as the nova, was used as the standard star for photometric 
observations. Aperture photometry was done using the APPHOT task in IRAF.

Spectral calibration was done using the OH sky lines that register with 
the stellar spectra. The spectra of the nearby field star SAO~174400 (1 Pup) 
were taken in $JHK$ bands at similar airmass as that of V574~Pup on all 
the observation nights  to ensure that the ratioing process (nova spectrum 
divided by the standard star spectrum) removes the telluric lines reliably. 
1~Pup, although an emission line star (spectral type A3 Iab), was chosen  
as the standard star due to its proximity to V574~Pup to minimise the 
effects of differential airmass between V574~Pup and the standard star. We 
have carefully removed the hydrogen lines in the spectra of 1~Pup (Pa$\beta$ 
and Br$\gamma$ were seen to be in emission; other Brackett lines are in
absorption) before taking the ratios. The ratioed spectra were then 
multiplied by a blackbody curve corresponding to the standard star's 
effective temperature to yield the final spectra. Extraction and reduction 
of the spectra were done using IRAF tasks.

\begin{table*}
\centering
\caption{Log of the Mt. Abu near-infrared observations of V574~Pup. The 
date of outburst is taken as 2004 November 20.67 UT.} 
\begin{tabular}{crrrrrrrccc}
\hline
\hline
Date of      &Days since    &\multicolumn{3}{|c|}{Integration time (s)} &\multicolumn{3}{|c|}{Integration time (s)} &\multicolumn{3}{|c|}{Nova Magnitude} \\
Observation  &outburst             &J-band      &H-band   &K-band  &J-band      &H-band   &K-band   &J-band      &H-band   &K-band  \\
\hline
   &   &\multicolumn{3}{|c|}{Spectroscopic Observations}   &\multicolumn{4}{|c|}{Photometric Observations}  \\
\hline
\hline
2004 Nov. 25	&5    &40  &40   &60   &28  &28  &24  &5.65$\pm$0.04 &5.32$\pm$0.09  &4.75$\pm$0.25 \\
2004 Nov. 26	&6    &60  &80   &120  &40  &20  &60  &6.11$\pm$0.04 &5.72$\pm$0.03  &5.39$\pm$0.03 \\
2004 Nov. 29	&9    &60  &60   &120  &36  &25  &12  &6.42$\pm$0.03 &6.09$\pm$0.02  &5.66$\pm$0.02 \\
2004 Dec. 01	&11   &60  &60   &60   &28  &16  &48  &6.31$\pm$0.12 &6.07$\pm$0.21  &5.73$\pm$0.14 \\
2004 Dec. 09	&19   &80  &60   &80   &64  &72  &116 &6.73$\pm$0.09 &7.46$\pm$0.33  &6.78$\pm$0.40 \\
2004 Dec. 14	&24   &60  &60   &80   &36  &36  &116 &7.18$\pm$0.05 &7.33$\pm$0.07  &6.86$\pm$0.40\\
2004 Dec. 24	&34   &100 &240  &180  &72  &72  &58  &7.89$\pm$0.02 &8.08$\pm$0.04  &7.49$\pm$0.07\\
2004 Dec. 25	&35   &180 &180  &180  &72  &72  &60  &7.92$\pm$0.04 &8.14$\pm$0.02  &7.34$\pm$0.22\\
2005 Jan. 01	&42   &120 &120  &240  &72  &72  &120 &8.25$\pm$0.02 &8.43$\pm$0.04 &7.71$\pm$0.04  \\
2005 Jan. 02	&43   &120 &120  &240  &72  &72  &120 &8.31$\pm$0.04 &8.53$\pm$0.04 &7.77$\pm$0.04\\
2005 Jan. 13	&54   &120  &120 &120  &144 &144 &116 &8.53$\pm$0.09 &8.61$\pm$0.09 &7.94$\pm$0.10\\
2005 Jan. 14	&55   &120  &120 &120  &----- &----- &-----  &--------  &-------- &--------\\
2005 Jan. 19	&60   &120  &120 &180  &----- &----- &-----  &--------  &-------- &--------\\
2005 Feb. 13	&85   &240  &240 &240  &360 &360 &200 &9.39$\pm$0.03 &9.44$\pm$0.03 &8.69$\pm$0.08 \\
2005 Feb. 19	&91   &-----  &----- &----- &400 &320 &200 &9.47$\pm$0.12 &9.48$\pm$0.14 &8.79$\pm$0.14   \\
2005 Feb. 25	&97   &360  &360 &360  &360 &360 &160  &9.48$\pm$0.07 &9.58$\pm$0.05 &8.58$\pm$0.14 \\
2005 Mar. 13	&113  &600  &600 &600  &540 &540 &198  &9.93$\pm$0.05  &10.05$\pm$0.11  &9.32$\pm$0.25   \\
2005 Mar. 19    &119  &-----  &----- &----- &480 &480 &198 &9.74$\pm$0.08 &9.88$\pm$0.11 &9.04$\pm$0.29 \\
2005 Mar. 20	&120  &600  &600 &600 &----- &----- &-----  &--------  &-------- &--------\\
\hline
\hline
\end{tabular}
\label{table1}
\end{table*}

\begin{figure}
\vskip 8.0 cm
\includegraphics{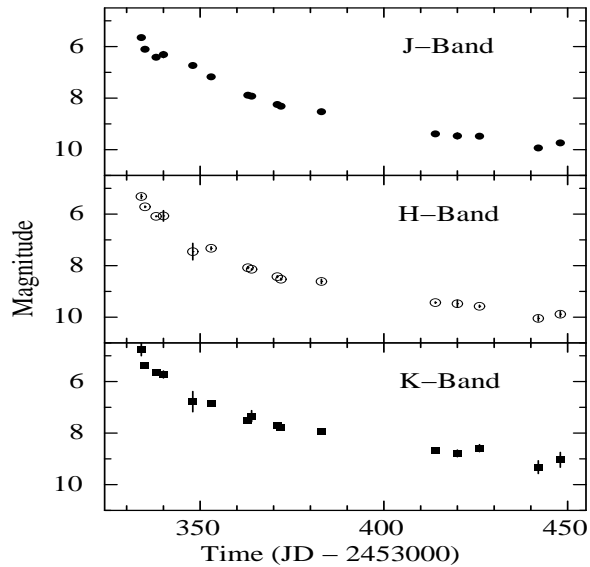}
\caption{The  light curves of V574~Pup in the $JHK$ bands obtained from the 
Mount Abu Observatory. The $J$, $H$ and $K$ band light curves are shown in
top, middle and bottom panels respectively.}
\end{figure}

\begin{figure*}
\includegraphics[bb=18 144 592 718, width=7.3in,height=6.5in,clip]{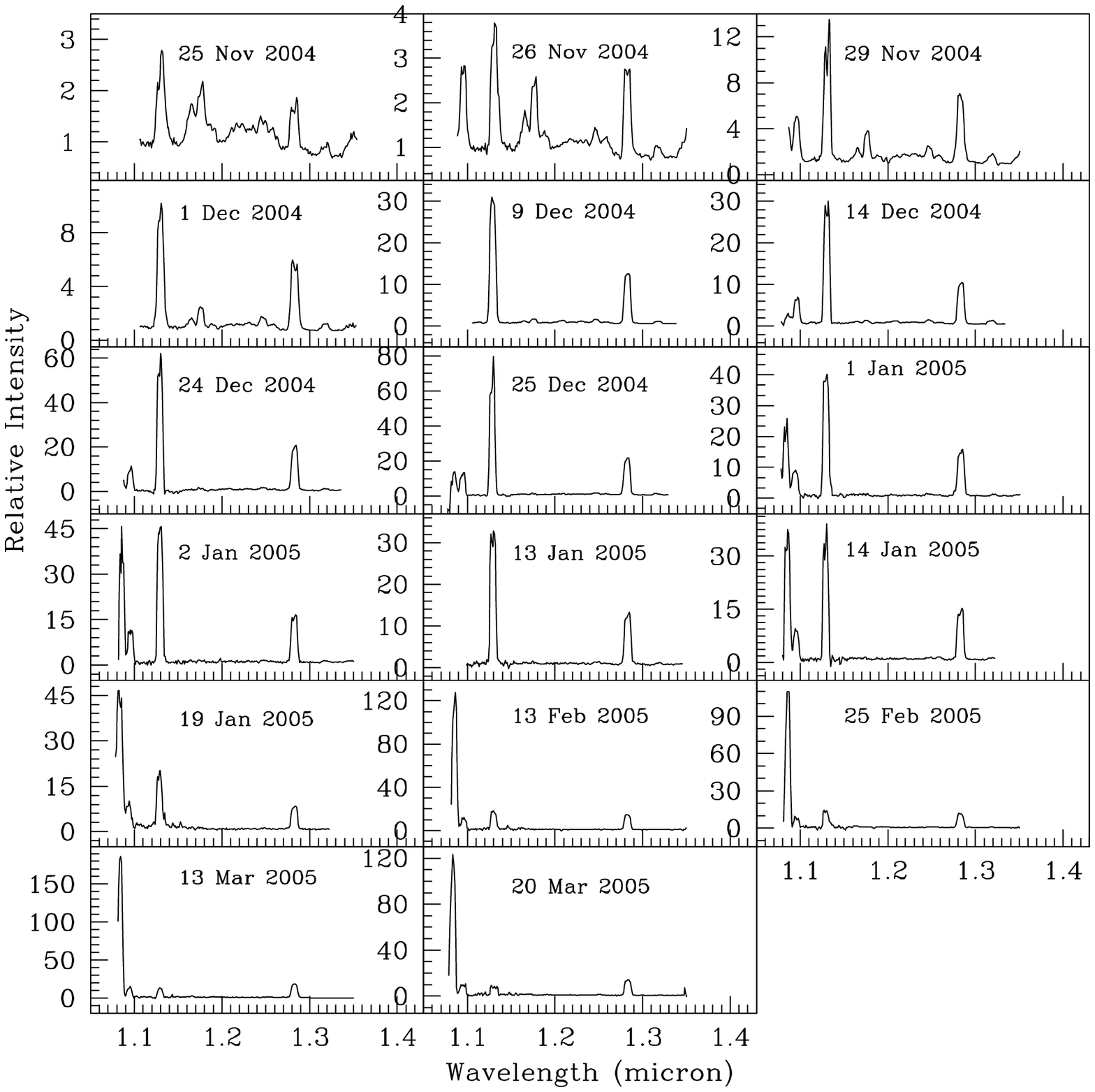}
\caption{The $J$-band spectra of V574~Pup at different epochs with the continuum  being normalized to unity at 1.25 \micron. }
\label{J-band}
\end{figure*}

\begin{figure*}
\includegraphics[bb=18 144 592 718, width=7.3in,height=6.5in,clip]{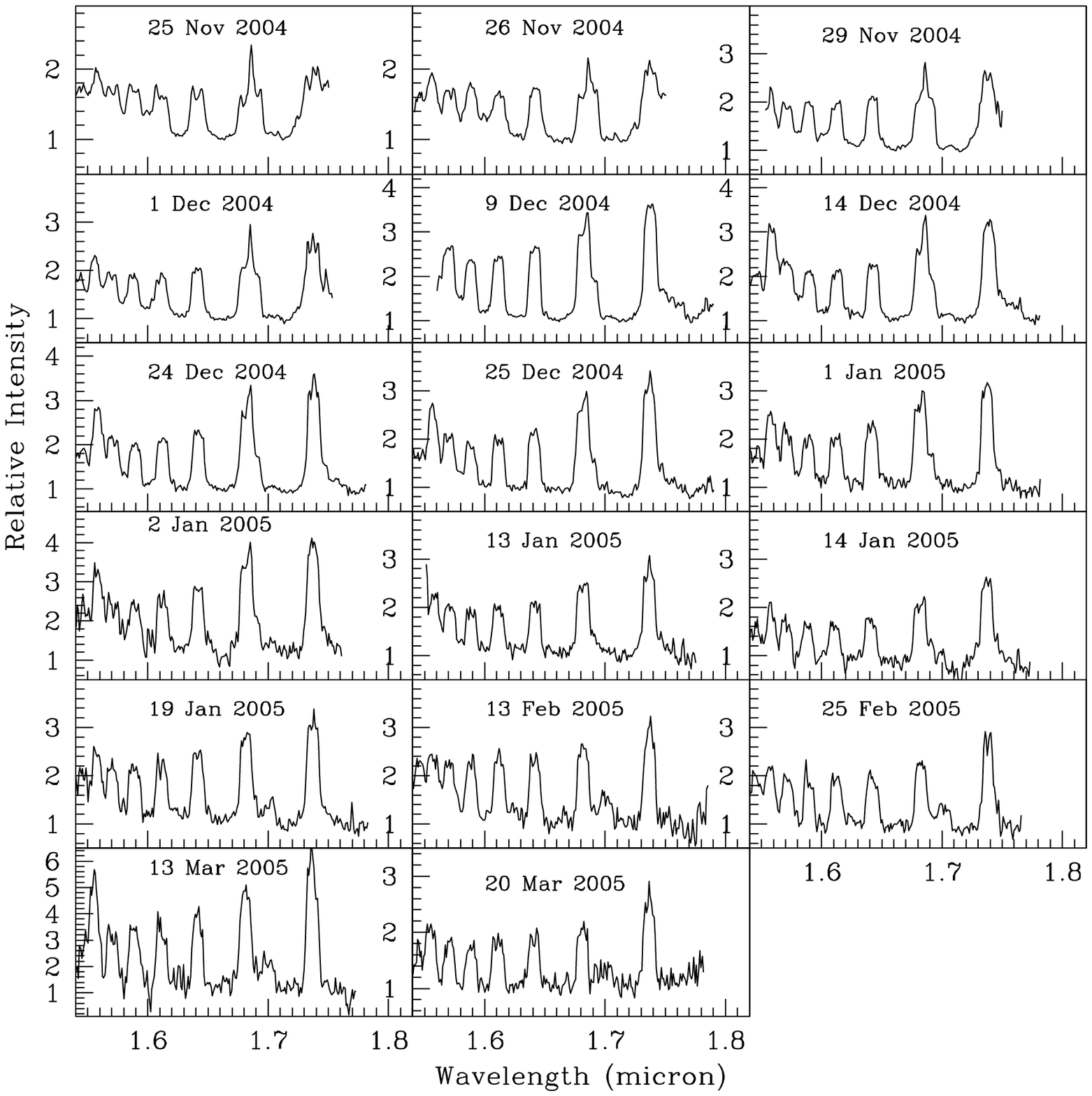}
\caption{The $H$-band spectra of V574~Pup at different epochs with the continuum  being normalized to unity at 1.65 \micron.}
\label{H-band}
\end{figure*}

\begin{figure*}
\includegraphics[bb=18 144 592 718, width=7.3in,height=6.5in,clip]{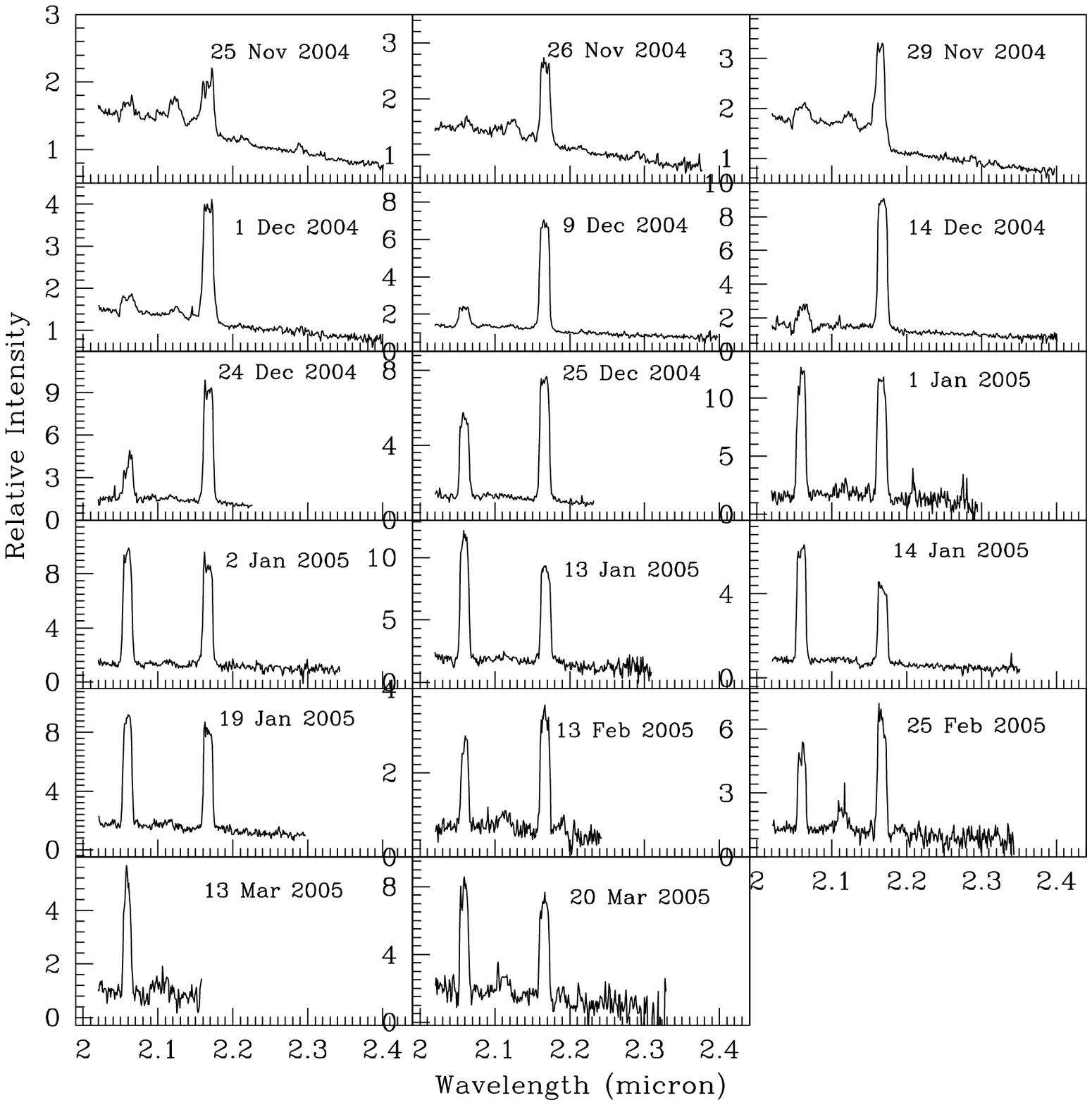}
\caption{The $K$-band spectra of V574~Pup at different epochs with the continuum  being normalized to unity at 2.2 \micron.}
\label{K-band}
\end{figure*}

\section{Results and Discussion}

\subsection{Distance estimation from the visual light curve}
We make a distance estimate to V574 Pup by analysing the $V$ band light 
curve using  archival data from the American Association of Variable Stars 
(AAVSO) and Association Francaise des Observateurs d’Etoiles Variables 
(AFOEV). The $V$ band light curve is presented in Figure~1 with 
AAVSO data shown in filled circles and AFOEV data shown in open circles. 
The present light curve shows that the discovery of V574 Pup took place 
on its way to the maximum that was reached on 2004 November 22.2611. The 
pre-maximum brightening lasted for one day and subsequent to the maximum 
there was a sharp drop by about 2 magnitudes again within a day. Following 
this drop V574 Pup showed a relatively slower rise reaching V=8.06 on 2004 
November 26.2458 and steadily decreased in brightness thereafter. For the 
purpose of calculation of $t_2$ and $t_3$, the time for a decline of 2 and 
3 magnitudes respectively, we assume that the sharp drop of 2 magnitudes 
lasting for a day can be ignored and take V$_{max}$ = 6.93 on 2004 
November 22.2611 UT (JD 2453331.7611). We then derive 
from the V-band light curve a value of  $t_2$=10$\pm$1 days and 
$t_3$=25$\pm$2 days. The absolute magnitude of the nova is then determined 
to be $M_V$=$-$8.73 using the maximum magnitude versus rate of decline 
(MMRD) relation of della Valle \& Livio (1995). At maximum (JD 2453331.7611)  
AAVSO lists the $B$ and $V$ magnitudes as 7.79 and 6.93 respectively. The 
mean intrinsic color of novae at maximum is estimated to be $(B-V)$$_0$ = 
+0.23$\pm$0.06 (Warner 1995). Using this relation, we obtain  $E(B-V)$ = 0.63 
and the extinction as $A_v$=1.95. We thus estimate  the distance to V574 Pup 
to be $d$=5.5 kpc and adopt this value in future calculations. Ness et al. 
(2007) have derived a distance of 3.2 kpc which is closer to the value  
obtained in the present analysis compared to the 15 - 20 kpc estimate by
Siviero et al. (2005). The significantly higher value for the distance 
estimated by Siviero et al. (2005) is due to the fainter value of $V_{max}$ 
considered by them and also their assumption of a negligible interstellar 
extinction towards the nova.

\subsection{$JHK$ light curves of V574~Pup}

The $JHK$ light curves of V574~Pup, obtained from the present photometric 
observations are presented in Figure~2. A gradual fading is seen 
in all the  three bands similar to the $V$ band behavior. Further, no rise 
is seen in any of the near-IR bands indicating the absence of  dust formation 
during the period of four months following the  outburst. We note from the 
figure, and also from the data in Table 1, that the $(J-H)$ color index is 
generally found to be negative - specially so during the later stages of 
the observations. A negative $(J-H)$ index is characteristic of novae as 
shown by Whitelock et al. (1984). The reason for this is the presence of 
strong emission features in the $J$ band spectra such as the Pa$\beta$ and 
Pa$\gamma$, HeI 1.083 \micron~ and the OI 1.1287 \micron~ lines. As 
discussed in the following subsection, the HeI 1.083 \micron~ and the 
OI 1.1287 \micron~ lines are specially strong in V574 Pup giving rise 
to the observed behavior of the $(J-H)$ color index (the  peak-to-continuum 
ratio of the HeI 1.083 \micron~ line goes as high as $\sim$ 150 in March 
2005). However, it may be noted that a negative $(J-H)$ index is not 
expected  if dust formation takes place (Whitelock et al. 1984; an example 
is the dust-forming nova V1280 Sco (Das et al. 2008)). The  observed $K$ 
band brightness of the nova is also  affected by significant contributions 
from the HeI  2.0585 \micron~ and Br$\gamma$ lines.

\subsection{Emission lines in the $JHK$ spectra}

The $JHK$ spectra of V574~Pup  are presented in Figures~3, 4 \& 5,
respectively. The early spectra cover 
the epoch of re-brightening seen at optical wavelengths and display 
typical emission lines of Paschen and Brackett series lines from hydrogen 
and the OI lines at 1.1287 \micron~ and 1.3164 \micron. Also seen 
prominently are  lines of carbon and nitrogen, particularly in the 
$J$ band. The details of the line identification are given in Table~2. 
The large observed ratio of the OI 1.1287 / 1.3164 \micron~ lines  
indicates that  Ly$\beta$ fluorescence is the dominant process 
contributing to the strength of the 1.1287 \micron~ line. The HeI 
line at 2.0581 \micron~ is clearly detected in the $K$ band spectra 
taken on November 29. This, and the $J$ band HeI line at 1.0830 \micron,  
gain in strength as the nova evolves. Their strength becomes comparable 
to hydrogen lines by 24-25 December and exceed them in strength by early 
January indicating a progress towards higher excitation conditions in the 
ejecta. Towards the end of our observations these HeI lines become very 
strong while other  weaker He~I lines at 1.7002 \micron~ and 2.1120-2.1132 
\micron~  are also  seen. We do not detect any coronal line features during 
the four months of our observational campaign. As mentioned earlier, many 
coronal features were detected in V574 Pup one year after the outburst that 
were seen to last for the next two years (Rudy et al. 2005, Lynch et al. 
2007).

During the course of our observations of V574~Pup, there were no 
significant changes seen in the width of the emission lines  in the $J$, 
$H$, \& $K$ band spectra. This implies the absence of any significant 
change in the expansion velocity of the ejected envelope. In order to 
quantify the changes in the line widths  in V574~Pup, the evolution of 
Brackett series lines which are prominent  was investigated. The overall 
width of the hydrogen Brackett series emission lines did not change 
appreciably during the observations. The mean velocity of the expanding 
gas, from the Br lines,  is estimated to be 1830$\pm$400 km s$^{-1}$ which 
agrees well with the findings by Rudy et al. (2006) who determined an 
average velocity of 1800 km s$^{-1}$.

It is also noted that no  lines from low ionization species like NaI or 
MgI are seen in the $JHK$ spectra.  These low ionization lines,
which are indicative of low temperature conditions, have been suggested as
potential diagnostic features to predict dust formation in the nova ejecta 
(Das et al. 2008). The absence of these lines in V574 Pup is consistent 
with the lack of dust formation in this nova. In this context, there 
is a line at 2.1452 \micron~ which matches a NaI transition at that wavelength. 
However, for reasons discussed in Das et al. (2008), we are doubtful whether this line should be attributed to NaI. A notable feature in the $H$ 
band spectra, is the  structure of the Br11 line at 1.6806 \micron~ which is
seen to be distinctly different from  other Brackett series lines in terms of 
both width and shape. This is most likely caused due to the blending of 
an additional nearby emission line of significant intensity.  In the 
following subsection, we attempt to identify this additional line and 
show that it most likely is due to a FeII line at 1.6872 \micron.

\subsection{Detection of the Fe~II 1.6872  \micron~ emission line in V574~Pup}

We examine the  distinct difference seen in the structure of Br11 line 
vis-a-vis other Br lines (Figure 4). This difference has persisted till 
the middle of 2005 January. The velocity of 3000$\pm$400 km s$^{-1}$ 
corresponding to the full width at half maximum (FWHM) of the Br11 
line is larger than the average value of 1830$\pm$400 km s$^{-1}$ 
corresponding to other Brackett series lines. This indicates that there 
is a definite contribution to Br11 line from an adjacent emission feature. 
It is noticed in the early spectra that  there is a central enhancement 
(spike) in the Br11 line which gradually decreases in strength and by late 
2004 December, Br11 line starts showing a prominent redward wing. Considering 
this behavior, we have looked for an emission line  on the higher wavelength 
side of Br11. In two  recent novae studied by us, namely, V2615~Oph (Das et 
al. 2009) and RS Oph (Banerjee et al. 2009) an emission line at 1.6872 
\micron~ has been clearly detected which is attributed to FeII. We suspect 
that this  FeII line is present here too. To study the effect of this line 
on  Br11 , we have generated synthetic line profiles by adding two lines 
viz. a primary line centered at 1.6806 \micron~ corresponding to Br11 and 
a second line at 1.6872 \micron~ corresponding to FeII. Since there is no 
apriori knowledge on the shape of the FeII line, we assume that its shape 
can be simulated by the the Br12 line profile (we choose Br12 because it 
is free from blending with other lines). We have similarly assumed that 
Br11 is well simulated by the  observed line profile of Br12 line. Keeping 
the peak intensity of synthetic Br11 line constant we have varied the 
intensity of the FeII line to simulate the observed temporal evolution. 
Two such synthetic spectra are shown in Figure~6. In the figure, 
we show the individual line profiles of Br11 and FeII with dotted lines 
and their resultant, co-added profile with a dashed line along with the 
observed profile (solid gray line). The left and right panels show the 
profiles for 2004 November 26 and 2004 December 9 respectively. The 
resultant and observed profiles are shown with some offset from the 
individual profiles for clarity.  It is seen that the synthetic profiles 
resulting from the combination of Br11 and Fe~II match  the observed line 
profile reasonably well. This indicates that  the  Fe~II 1.6872 \micron~ 
line is present in V574 Pup. Based on the work of Banerjee et al. (2009), 
another Fe II line at 1.7413 \micron~ could also be expected in the 
spectrum. It is possible that this line is also there, but it is difficult 
to draw a definitive conclusion regarding its presence since it could be 
blended, rather too closely with Br10 and  also a cluster of CI lines in 
the 1.74-1.77 \micron~ region. We note that these FeII lines are not too 
commonly reported in the spectra of novae. Apart from V2615 Oph (Das et 
al. 2009) and RS Oph (Banerjee et al. 2009), there are two more novae, 
namely V2540 Ophiuchi (Rudy et al. 2002) and CI Aquila (Lynch et al. 
2004) where these lines appear to be detected. The excitation mechanism 
for these lines is believed to be Lyman $\alpha$ and Lyman continuum 
fluorescence coupled with collisional excitation (Banerjee et al. 2009 
and references therein).

\begin{figure}
\includegraphics[bb=0 0 301 292, width=3.5in,height=3.5in,clip]{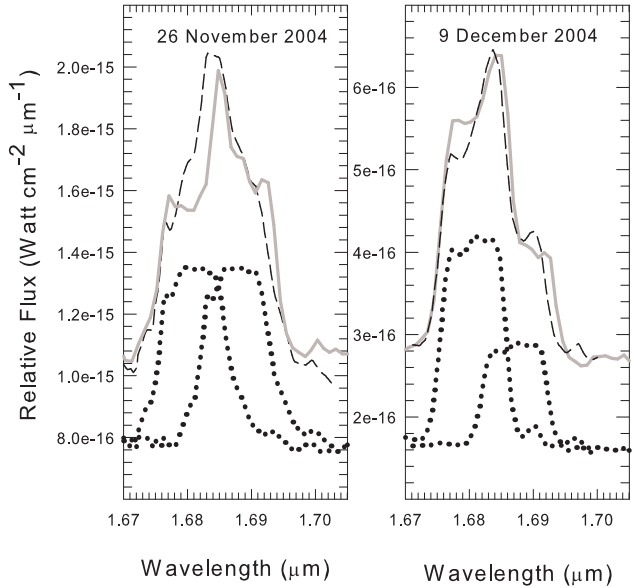}
\caption{The profiles of Br11 line at 1.6806 \micron~ (dotted line), Fe~II line at 1.6872 \micron~ (dotted line), the added line (Br11+Fe~II) intensity (dashed line) and the observed data (gray solid line) are shown for 2004 November 26 (left panel) and for 2004 December 9 (right panel). The intensity of the Fe~II line was adjusted to match the co-added profile with the observed profile. The co-added and observed profiles are shown with some offset in order to make it clear for comparison.}
\label{feII}
\end{figure}

\begin{table}
\begin{center}
\caption{List of prominent lines in the $JHK$ spectra}
\begin{tabular}{l l r }
\hline
\hline
Wavelength        &Species   & Remarks \\
(${\rm{\mu}}$m)   &          & \\
\hline 
\hline  
1.0830   &He I            \\
1.0938   &Pa $\gamma$     \\
1.1287   &O~I                 \\
1.1600 - 1.1674	&C~I	&strongest lines at \\
                &       &1.1653, 1.1659, 1.16696 \micron \\
1.1748 - 1.1800	&C~I	&strongest lines at \\
                &       &1.1748, 1.1753, 1.1755 \micron \\
1.1819 - 1.1896	&C~I	&strongest lines at \\
                &       &1.1880, 1.1896 \micron \\
1.2187 - 1.2382	&C~I, N~I& blend of N~I 1.2187, \\
& & 1.2204, 1.2329, 1.2382, \\
& & \& CI 1.2249, 1.2264 \micron \\
1.2461, 1.2469  &N~I    \\
1.2562 - 1.2614   &C~I    & blend of C~I 1.2562, 1.2569\\
& & 1.2601, 1.2614 \micron \\
1.2818          &Pa $\beta$    \\
1.3164          &O~I           \\
1.5439	        &Br 17		\\
1.5557	 	&Br 16         \\
1.5685	 	&Br 15         \\
1.5881	 	&Br 14         \\
1.6005		&C~I	&may be present\\
1.6109  	&Br 13         \\
1.6407   	&Br 12         \\
1.6806   	&Br 11         \\
1.6872		&Fe~II		\\
1.7002          &He~I           \\
1.7045   	&C~I		\\
1.7362   	&Br 10         \\
1.74 - 1.77       &CI    &blend of several C~I Lines\\
2.0585	   	&He I          \\
2.1120, 2.1132   &He~I            \\
2.1156 - 2.1295	&C~I	&blend of several C~I\\
		&	&lines, strongest being\\
		&	&2.1156, 2.1191, 2.1211,\\
		&	&2.1260, 2.1295 \micron\\
2.1452		&Na~I?	\\
2.1655   &Br $\gamma$   \\
2.2156-2.2167	&C~I	&blend of C~I lines at\\
		&	&2.2156, 2.2160, 2.2167 \micron\\
2.2906		&C~I\\
\hline
\hline
\end{tabular} 
\\
\end{center}
\label{line-info}
\end{table}

\subsection{Evolution of the continuum}
We analyse and discuss the  evolution of the continuum spectra of V574~Pup 
here. At the time of outburst, a nova's continuum is generally well 
described by a blackbody distribution from an optically thick 
pseudo-photosphere corresponding to a stellar spectral type A to F 
(Gehrz 1988). The spectral energy distribution then gradually evolves 
into a free-free continuum as the optical depth of the nova ejecta 
decreases (Ennis et al. 1977; Gehrz 1988).  The evolution of the continuum
of V574~Pup is shown in Figure 7 wherein we have shown representative 
spectra sampling the duration of our observations. The spectra in Figure~7 
were flux calibrated using the broadband $JHK$ photometric observations 
presented in Table 1. During this process of calibrating  the flux in the 
continuum, we note that the observed broadband flux is a sum of 
contributions from both the continuum and also from emission lines.  
As the emission lines of HI, OI and HeI are strong and contribute 
significantly to the broadband fluxes, we have first calculated the 
contribution of all the prominent emission lines to the observed 
spectra and removed this contribution from the broad band photometric 
fluxes measured from the $JHK$ photometry. This gives the true continuum 
flux which was used to calibrate the spectra in Figure~7.

We have tried to fit  the spectra in Figure~7 with power law fits
i.e. $F_\lambda~\propto~\lambda^{-\alpha}$. In the beginning of our 
observation i.e. on 2004 November 25 (4 days after outburst) and 2004 
December 01, the continuum spectrum approximately fits  a spectral 
index of $\alpha$ $\sim$ 2.75. A blackbody fit, expected to have an 
index of $\alpha$ = 4.0 at longer wavelengths, does not simulate the 
data too well. The subsequent spectra  gradually become flatter with 
a slope of $\alpha$ $\sim$ 2.0 on 2004 December 14. The nova continuum 
is subsequently found to become flatter with time and  gradually match 
a free-free emission spectrum towards the end of our observation campaign. 
The spectrum of  2005 February 25 in Figure 7 has been fit with a free-free 
spectrum computed at a temperature of T = 10$^4$ K. 

\begin{figure*}
\centering
\includegraphics[bb= 0 0 613 793, width=6.1in,height=6.5in,clip]{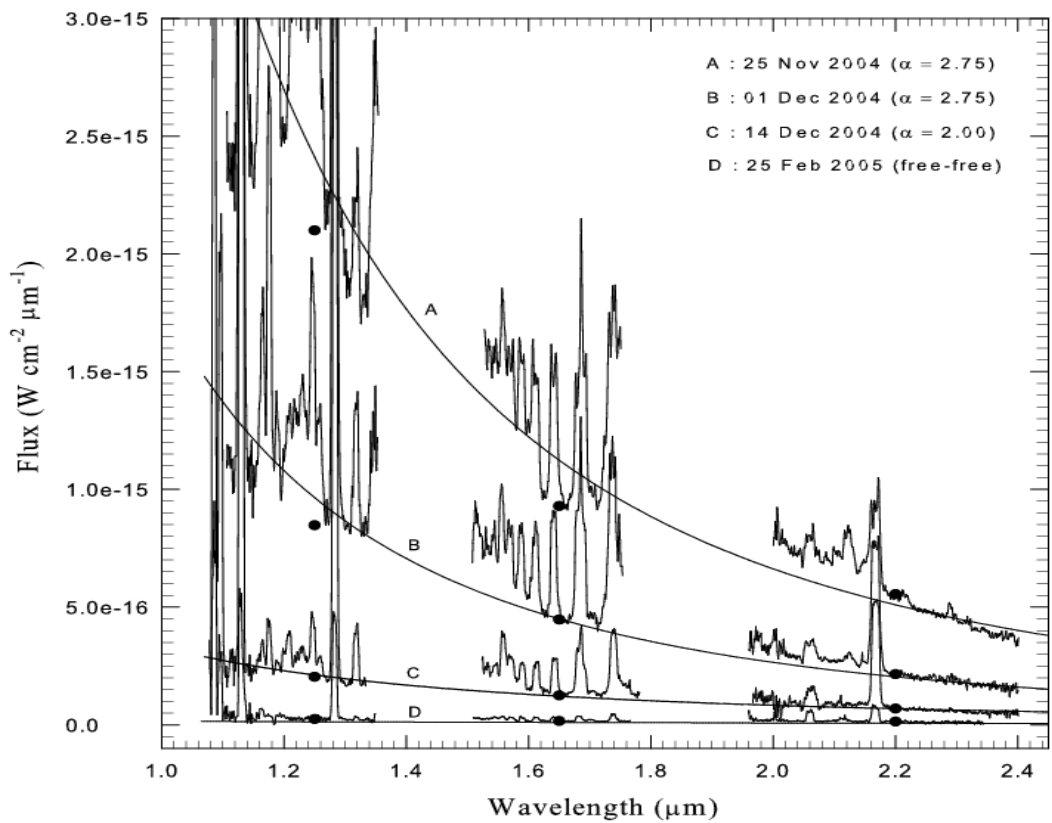}
\caption{The composite $JHK$ spectra of V574~Pup for 25 November 2004 (A), 
1 December 2004 (B), 14 December 2004 (C), and 25 February 2005 (D) from 
near-infrared observations with the Mt Abu telescope. Model fits to the 
data (using either a power law or a free-free spectral dependence) are 
shown by the continuous lines;  the broadband fluxes (corrected for 
contribution from line emission - see text) are shown by filled circles. 
The slope of the continuum spectra of A, B, and C are compared with power 
law fits  ($F_\lambda~~\propto~~\lambda^{-\alpha}$) with slopes $\alpha$ 
= 2.75, 2.75 and 2.0, respectively. A free-free emission function at 
temperature 10$^4$ K is plotted along with the fourth spectrum (noted 
as D). It appears that the nova continuum has flattened gradually to a 
free-free type of emission towards the end of our observations.}
\end{figure*}

\subsection{Case~B recombination analysis}
The recombination case~B analysis for the HI lines were carried out for 
all the observed spectra and the representative results for five epochs 
covering the first 80 days of our observations are shown in 
Figure~8. We have plotted in Figure~8 the observed 
relative strength of Brackett series lines with the line strength of Br12 
as unity along with the predicted values for three different recombination 
case~B emissivity values from Storey \& Hummer (1995). These predicted 
values cover  a representative temperature of T = 10$^4$ K and the electron 
densities of $n_e$ = 10$^9$ cm$^{-3}$, 10$^{10}$ cm$^{-3}$ and 10$^{12}$ 
cm$^{-3}$. High electron densities are considered because the ejecta 
material is expected to be dense in the early stages after the outburst. 
For the early epochs,namely 2004 November 25 and 2004 December 1, the Br10 
line is not included as it is at the edge of the observed spectra and not 
adequately covered. The errors in the estimated line strengths  are 
$\sim$10\% for the Br$\gamma$, Br12 $\&$ Br13 lines and   $\sim$20\% for 
the Br14,  Br15, Br16 \& Br17 lines. The errors for the Br11 line are 
$\sim$30\% for 2004 November 25, $\sim$20\% for 2004 December 1 and 
$\sim$10\% for observations on the other days. The variable error 
assigned to the Br11 is due to the presence of a Fe~II line at 1.6872 
\micron~ that was strong in the initial phase of our observations, 
gradually  weakened and finally became undetectable.

Figure~8 shows that the observed line intensities clearly deviate 
from  case~B values in the initial phase of our observations. Specifically, 
Br $\gamma$ which is expected to be the relatively stronger than the other 
Br lines, is observed to be considerably weaker in the early observations. 
This is most likely  due to optical depth effects  in the Brackett lines 
(Lynch et al 2000). Such deviations from the recombination case~B conditions 
during the early stages after outburst can be expected and have been observed 
in other novae too, for example, V2491~Cyg and V597~Pup (Naik et al. 2009), 
RS~Oph (Banerjee et al. 2009) etc. However, towards the end of the 
observations, on 2005 February 13 and 25 (fourth and fifth panels of 
Figure~8) there is an indication that Case B conditions have 
begun to prevail. For these last two dates, it is found that the observed 
data matches well with the predicted values for the recombination case~B 
values of T = 10$^4$ K and an electron density in the range  $n_e$ = 
10$^{9-10}$ cm$^{-3}$.

\begin{figure}
\begin{center}
\includegraphics[width=12.0cm,angle=-90]{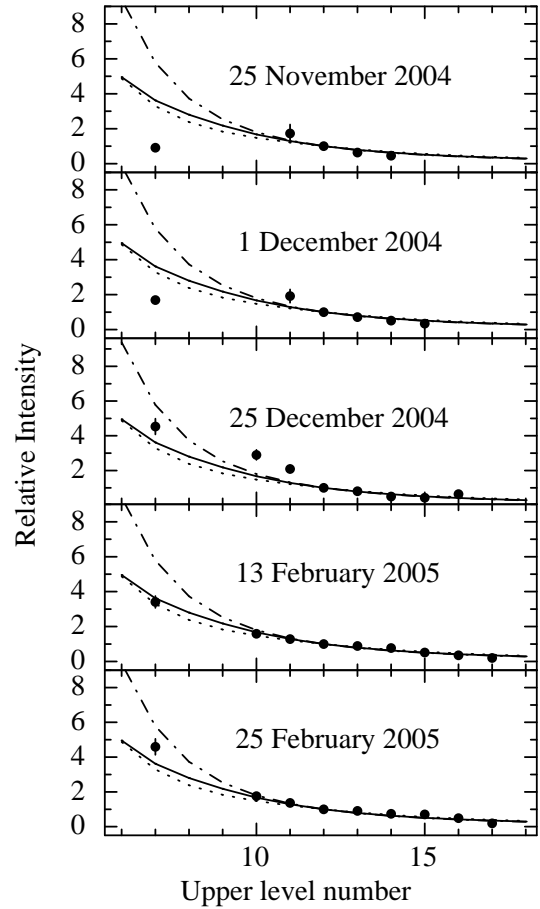}
\caption{Recombination analysis for the hydrogen Brackett lines in V574~Pup 
on selected dates of our near-IR observations (as noted in the figure). The 
abscissa is the upper level number of Brackett series line transition. The 
line intensities are relative to that of Br~12 which is normalized to unity. 
The errors in the estimated line strengths  are $\sim$10\% for the Br$\gamma$, 
Br12 $\&$ Br13 lines and   $\sim$20\% for the Br14,  Br15, Br16 and Br17 lines.
The errors for the Br11 line are $\sim$30\% for 2004 November 25, $\sim$20\% 
for 2004 December 1 and $\sim$10\% for observations on the other days. The 
case~B model predictions for the line strengths are also shown for a 
temperature $T=10^4 K$ and electron densities of $n_e = 10^{12} cm^{-3}$ 
(dot-dash line), 10$^{10} cm^{-3}$ (solid line), and 10$^{9} cm^{-3}$ 
(dotted line).}  
\end{center}
\end{figure}

\subsection{Estimation of the mass of the nova ejecta}
 
We  estimate the mass of the ionized gas in the ejecta  using the 
fact that on 2005 February 25, the observed SED of V574 Pup is well 
fit by a free-free flux distribution at a temperature of T = 10000K 
(Figure~7). 

The free-free volume emission coefficient from an ionized gas is given by \\
[-3mm]

$j$${_{\rm \lambda ff}}$ $=$ 2.05$\times$10${^{\rm -30}}$~$\lambda$${^{\rm -2}}$
$z$${^{\rm 2}}$ {\rm{$g$}} $T$${^{\rm -1/2}}$ $n$${_{\rm e}}$~$n$${_{\rm i}}$ 
$e$${^{(\rm -c{_{2}}/\lambda \rm T)}}$  \\ [-6.5mm]
          \begin{flushright}
          W cm${^{\rm -3}}$ ${\rm{\mu}}$m${^{\rm -1}}$ \\
          \end{flushright}
          \vskip -1mm
          \noindent

where $\lambda$ is the wavelength in \micron, $z$ is the charge, $g$ is the 
Gaunt factor (assumed equal to unity), $T$ is the temperature, n$_e$ and 
n$_i$ are the electron and ion densities respectively and c$_2$ = 1.438 cm K. 
The total continuum emission from the nova ejecta can then be estimated by 
multiplying the flux given in the  above equation with the shell volume 
$V$${_{\rm s}}$ and equating it  to the observed flux F$_{\lambda ff}$ which 
equals:  

$F$${_{\rm \lambda ff}}$ $=$ $j$${_{\rm \lambda ff}}$~$\times$~$V$${_{\rm s}}$/4${\rm{\pi}}$$d$${^{\rm 2}}$ 

where $d$ is the distance to the object. We use $d$ = 5.5 kpc, T = 10$^4$K 
and  n$_e$ = n$_i$ assuming a pure and completely ionized hydrogen ejecta 
($z$ = 1). At $K$ band center ($\lambda$ = 2.2 \micron), using the observed 
flux on 2005 February 25 to be 1.325$\times$10$^{-17}$ W cm$^{-2}$ 
\micron$^{-1}$ and n$_e$ = 10$^{10}$ cm$^{-3}$ derived from the 
recombination analysis, we obtain the volume of the emitting HI gas 
to be $V$${_{\rm s}}$ = 2.2$\times$10$^{41}$ cm$^3$. Similar values 
for $V$${_{\rm s}}$ are obtained if the $J$ and $H$ band observed 
fluxes and corresponding central wavelengths of these bands are used 
instead. 

The mass of the ionized gas can then be calculated using M$_{gas}$ = 
$V$${_{\rm s}}$$n_e$$m_H$, where $m_H$ is mass of the hydrogen atom. 
Taking n$_e$=10$^{10}$ cm$^{-3}$  gives a value of M$_{gas}$ = 
1.8$\times$10$^{-6}$ M$_\odot$. A similar calculation for n$_e$=10$^{9}$ 
cm$^{-3}$, which could also be a valid estimate in V574~Pup (as indicated 
from case~B analysis for 25 Feb 2005), yields  M$_{gas}$ = 
1.8$\times$10$^{-5}$ M$_\odot$. Thus, within the uncertainties 
associated with the parameters involved, we would estimate the 
mass of the ionized gas to be in the range of 1.8$\times$10$^{-5 }$ 
to 1.8$\times$10$^{-6 }$M$_\odot$. This estimate is reasonably in 
agreement with the  observed range of  the mass of  novae ejecta 
(1 to 30 $\times$ 10$^{-5}$ M$_\odot$) and also the theoretically 
calculated range of 5.3$\times$10$^{-8}$ to 6.6$\times$10$^{-4}$ 
M$_\odot$ in the extended grid of models computed by Yaron et al. 
(2005).

The mass estimate made above can be checked for consistency through 
an alternative approach. As discussed in section 3.6,  on 13 and 25 
February 2005 there is a reasonably good match between the observed 
line intensities of HI Brackett series lines with the theoretical case~B 
values listed in Storey and Hummer (1995) for T=10$^4$ K and 
$n_e = 10^{9-10} cm^{-3}$. For these values, the emissivity in 
Br $\gamma$ is expected to be $j(Br\gamma)$ $\sim$ 4.5$\times$10$^{-14}$ 
W cm$^{-3}$ (Storey and Hummer, 1995). From the observed data of 2005 
February 13 and 25, the Br $\gamma$ line is measured to have a mean 
line luminosity of $\sim$ 5.0$\times$10$^{-19}$ W cm$^{-2}$. Using 
$d$ = 5.5 kpc, the total power in the line is thus  1.8$\times$10$^{28}$ W. 
With $j(Br\gamma)$ as estimated above, this yields the volume of the 
emitting gas to be $V$${_{\rm s}}$ = 4.0$\times$10$^{41}$ cm$^3$ 
(consistent with the value of 2.2$\times$10$^{41}$ cm$^3$ derived 
from the free-free analysis) which thereby leads to a similar mass 
estimate as made earlier. 

\section{Summary}
We have presented an extensive set of spectroscopic and photometric 
observations of nova V574~Puppis. From the $V$ band light curve  the 
distance to the nova  is estimated to be $\sim$5.5 kpc. The near-IR 
light curve  shows a steady decline with time without any evidence 
for the buildup of an infrared excess associated with dust formation 
in the ejecta. However, an infrared excess at longer wavelengths 
can not be ruled out. Along with lines of  hydrogen, helium, oxygen and 
carbon, we also detect the  Fe~II emission line at 1.6872 \micron~ 
in the near-IR spectra. The nova continuum is modeled and found to 
evolve from a $\lambda^{-2.75}$ dependence to a free-free  emission 
during the period of our observations. A recombination analysis of 
the HI lines is presented.  We estimate the mass of the ionized gas 
in the ejecta and show it to lie in the range 10$^{-5}$ M$_\odot$ and 
10$^{-6} $M$_\odot$.

\section*{Acknowledgments}
We wish to thank the referee Prof. A Evans for his useful suggestions 
on the paper. The research work at Physical Research Laboratory is 
funded by the Department of Space, Government of India. We thank George 
Koshy for help with some of the observations. We acknowledge with thanks 
the variable star observations from the AAVSO International Database, 
contributed by observers worldwide, and used in this research. This 
research has made use of the AFOEV database, operated at CDS, France.


\begin{thebibliography}{99}
\bibitem[]{}Ashok, N. M., Banerjee, D. P. K., 2004, IAU Circ., 8447, 4
\bibitem[]{}Ayani, K. 2004, IAU Circ., 8443, 2
\bibitem[]{}Banerjee, D. P. K., Das, R. K., Ashok, N. M., 2009, MNRAS, 399, 357
\bibitem[]{}Das, R. K., Banerjee, D. P. K., Ashok, N. M., Chesneau, O., 2008,
            MNRAS, 391, 1874
\bibitem[]{}Das, R. K., Banerjee, D. P. K., Ashok, N. M., 2009, MNRAS, 398, 375
\bibitem[]{}della Valle, M., Livio, M., 1995, ApJ, 452, 704
\bibitem[]{}Ennis, D., Beckwith, S., Gatley, I., Matthews, K., Becklin, E. E., Elias, J., Neugebauer, G., Willner, S. P., 1977, ApJ, 214, 478 
\bibitem[]{}Gehrz, R. D., 1988, ARA\&A, 26, 377
\bibitem[]{} Lynch D.K., Rudy R.J., Mazuk S., Puetter R.C., 2000, ApJ, 541, 791 
\bibitem[]{}Lynch, D. K., Wilson, J. C., Rudy, R. J., Venturini, C., Mazuk, S., Miller, N. A., Puetter, R. C., 2004, AJ, 127, 1089
\bibitem[]{}Lynch, D. K., Rudy, R. J., Prater, T. R., Gilbert, A. M., Mazuk, S., Perry, R. B., Puetter, R. C., 2007, IAU Circ., 8906, 1
\bibitem[]{}Naik, S., Banerjee, D. P. K., Ashok, N. M., 2009, MNRAS, 394, 1551
\bibitem[]{}Nakano, S., Tago, A., Sakurai, Y., Kushida, R.,  Kadota, K., 2004, IAU Circ., 8443, 1
\bibitem[]{}Ness, J.-U., Schwarz, G. J., Retter, A., Starrfield, S., Schmitt, J. H. M. M., Gehrels, N., Burrows, D., Osborne, J. P., 2007, ApJ, 663, 505
\bibitem[]{}Rudy, R. J., Lynch, D. K., Mazuk, S., Venturini, C. C., Puetter, R. C., Perry, R. B., 2002, AAS, 201, 4006
\bibitem[]{}Rudy, R. J., Lynch, D. K., Mazuk, S., Venturini, C. C., Puetter, R. C., Perry, R. B., Walp, B., 2005, IAU Circ., 8643, 2
\bibitem[]{}Rudy, R. J. et al., 2006, AAS, 209, 906
\bibitem[]{}Samus, N. N., Kazarovets, E., 2004, IAU Circ., 8445, 2
\bibitem[]{}Siviero, A., Munari, U., Jones, A. F., 2005, IBVS, 5638, 1
\bibitem[]{}Storey, P. J., Hummer, D. G., 1995, MNRAS, 292, 41
\bibitem[]{}Warner, B., 1995, {\it Cataclysmic Variable Stars. Cambridge Astrophysics Series}, Cambridge Univ. Press, Cambridge, New York 
\bibitem[]{}Whitelock, P. A., Carter, B. S., Feast, M. W., Glass, I. S., Laney, D., Menzies, J. W., Walsh, J., Williams, P. M., 1984, MNRAS, 211, 421
\bibitem[]{}Yaron, O., Prialnik, D., Shara, M. M., Kovetz, A., 2005, ApJ, 623, 398
\end{thebibliography}
\end{document}